\documentclass[secnumarabic, amssymb, nobibnotes,twocolumn,aps]{revtex4}
\usepackage{color}

\usepackage{graphicx}

\begin{document}

\title{Relation of open circuit voltage to charge carrier density in organic bulk heterojunction solar cells}

\author{Daniel Rauh$^1$}
\email[Electronic mail: ]{daniel.rauh@zae.uni-wuerzburg.de}
\author{Alexander Wagenpfahl$^2$}
\author{Carsten Deibel $^2$}
\author{Vladimir Dyakonov$^{1,2}$}
\email[Electronic mail: ]{dyakonov@physik.uni-wuerzburg.de}
\affiliation{$^1$ ZAE Bayern, Bavarian Center for Applied Energy Research, Am Hubland, 97074 W\"urzburg, Germany}
\affiliation{$^2$ Experimental Physics VI, Faculty of Physics and Astronomy, Julius-Maximilians-University of W\"urzburg, Am Hubland, 97074 W\"urzburg, Germany}

\begin{abstract}
The open circuit voltage $V_{oc}$ and the corresponding charge carrier density were measured in dependence of temperature and illumination intensity by current--voltage and charge extraction measurements for P3HT:PCBM and P3HT:bisPCBM solar cells. At lower temperatures a saturation of $V_{oc}$ was observed which can be explained by energetic barriers at the contacts (metal-insulator-metal model). Such injection barriers can also influence $V_{oc}$ at room temperature and limit the performance of the working solar cell, as was assured by  macroscopic device simulations on temperature--dependent IV characteristics. However, under most conditions --  room temperature and low barriers -- $V_{oc}$ is given by the effective bandgap.  
\end{abstract}

\maketitle

The open circuit voltage is one of the key parameters to optimize organic bulk heterojunction (BHJ) solar cells and therefore under intensive investigation \cite{deibel2010b}. Recent publications already showed a linear dependence of $V_{oc}$ on the enery difference between the acceptor LUMO and polymer HOMO \cite{brabec2001}. This effective band gap $E_g$ was later attributed to the energy of the charge transfer state \cite{vandewal2010,deibel2010} and is determining the maximum value of the magnitude $eV_{oc}$, (e is elementary charge) that can be achieved in a particular system. This upper limit is reduced by surface \cite{ramsdale2002} and bulk recombination \cite{vandewal2010} by 0.3 to 0.5~$eV$ \cite{scharber2006, vandewal2008}. Choosing the right electrode material can minimize the influence of the surface losses. 

Koster \textit{et al.} derived an analytical equation for $V_{oc}$ based on Shockley, drift and continuity equation under assumption of Langevin recombination resulting in 
\begin{equation}
V_{oc}  =  \frac{E_g}{q} - \frac{k_B T}{q}\ln\left(\frac{\left(1-P\right)k N_c^2}{P G }\right) \\
\label{eq:voc_bilayer}
\end{equation}
with  the temperature $T$, the Boltzmann constant $k_B$, the elementary charge $q$, the Langevin recombination constant k, the effective density of states $N_c$, the generation rate of bound polaron pairs $G$ and the dissociation rate of these into free polarons $P$ \cite{koster2005}. This equation is equivalent to 
\begin{equation}
V_{oc}   =  \frac{E_g}{q} - \frac{k_B T}{q}\ln\left(\frac{N_{c}^2}{n\cdot p}\right).
\label{eq:voc2}
\end{equation}
with the charge carrier densities of the electrons and holes in the device $n$ and $p$ \cite{cheyns2008}. Under the assumption of a constant $N_c$ for a specific donor--acceptor blend, a higher steady-state charge carrier concentration leads to an increase of $V_{oc}$. There are two ways to get a higher equilibrium charge carrier density in the device at open circuit conditions: a higher generation rate of polarons and a lower recombination rate. Internal quantum efficiency values close to 100~\% were reported for P3HT:PCBM \cite{schilinsky2002} at short circuit conditions. In addition, photocurrent measurements on the same material system showed, that the polaron pair dissociation yield $P$ at room temperature has a weak voltage dependence in the range of short circuit to open circuit  \cite{limpinsel2010,deibel2009}. Therefore the gain in carrier concentration at a constant light illumination by increasing the generation rate of these and thus $V_{oc}$ is very limited. This leads to the second and more promising point: the influence of nongeminate recombination. Reducing the recombination rate will increase the steady state polaron concentration and thus  the open circuit voltage.

In this article the open circuit voltage is investigated temperature and illumination dependent together with the corresponding $n$ in the device. We show that, depending on temperature, the $V_{oc}$ is mainly determined by the effective bandgap of the donor--acceptor blend or by the metal--insulator--metal (MIM) model. 

Three different types of solar cells were processed. Two devices had an active layer consisting of poly(3-hexylthiophene-2,5-diyl) (P3HT, P200 from Rieke Metals) as donor blended with [6,6]-phenyl-C$_{61}$butyric acid methyl ester (PCBM, from Solenne) as acceptor in the ratio of 1:0.8. For one cell a Ca (3~nm) / Al (100~nm) cathode was used, for the other a Cr (3~nm) / Al (100~nm) cathode known to form an injection barrier . For the third solar cell, bis(1-[3-(methoxycarbonyl)propyl]-1-phenyl)-[6.6]C62 (bisPCBM, from Solenne) was used as acceptor in a ratio of 1:1 with P3HT and a Ca (3~nm) / Al (100~nm) contact. The cells were processed as follows.  Structured indium tin oxide (ITO)/glass was cleaned successively in soap water, acetone and isopropanol for at least 10~min in an ultrasonic bath before a thin layer of poly(3,4-ethylendioxythiophene):polystyrolsulfonate (PEDOT:PSS, CLEVIOS P VP AI 4083) was spincoated on ITO to serve as anode. After transferring the samples into a nitrogen filled glovebox  a heating step of 130 $^\circ C$ for 10~min was applied. The active layers were spincoated from chlorobenzene solutions to gain an active layer thickness for all three cells of about 200~nm. The samples were annealed again for 10~min at 130~$^\circ C$ before the metal contacts were thermally evaporated at a pressure below $1\cdot 10^{-6}$~mbar. 

All temperature dependent current--voltage (IV) characteristics and charge extraction (CE) measurements  were performed in a cryostate. As light source a 10~W  white LED was used. IV-curves were recorded to extract $V_{oc}$. $n$ was determined by the CE technique (for detail see Ref.~\cite{shuttle2008b}). The used CE setup consisted of a function generator for applying $V_{oc}$ to the solar cell and triggering the LED. The CE signal was preamplified before it was detected with an oscilloscope. Integrating the obtained signal resulted in the number of extracted charges at $V_{oc}$. This value was then corrected by capacitance effects. The recombination of charge carriers during the extraction process was determined to be low (a few \%) and therefore not taken into account \cite{shuttle2008b}.  The charge carrier density was calculated by considering the volume of the active layer.

\begin{figure}[tbp] 
   \centering
   \includegraphics[scale=0.6]{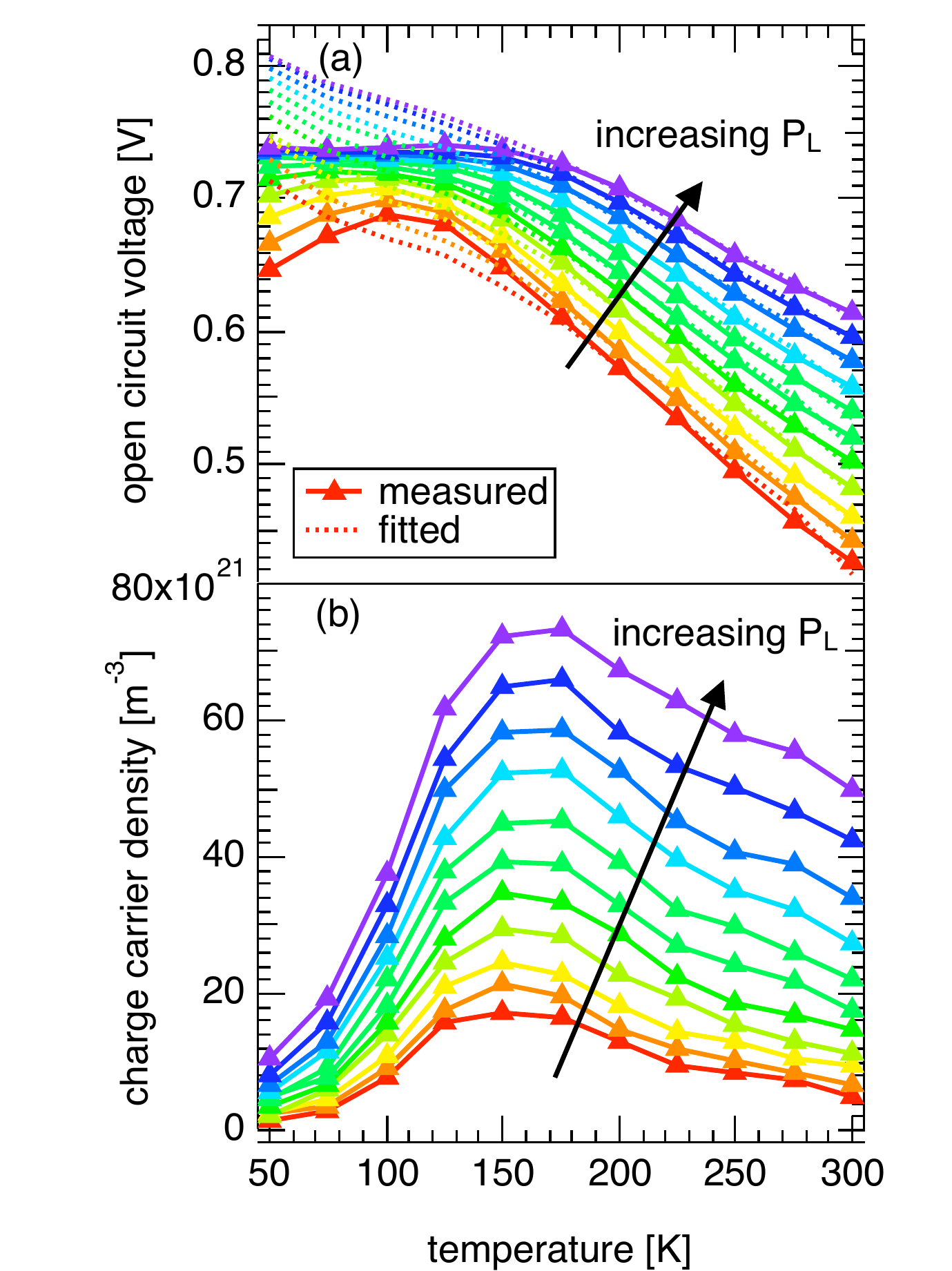} 
   \caption{(Color online) Temperature dependent $V_{oc}$ (a) and corresponding charge carrier density (b) for different illumination levels ranging from 0.01 - 3.16~suns of an annealed P3HT:PCBM solar cell. The dotted lines in (a) indicate the fit using Eq.~(\ref{eq:voc2}) as fit function.}
   \label{fig:Fig1}
\end{figure}
In Fig.~\ref{fig:Fig1} temperature and illumination  dependent $V_{oc}$ and the corresponding charge carrier densities $n$ are shown for the P3HT:PCBM solar cell with Ca/Al contacts. 
Similar dependencies of $V_{oc}$ have been previously reported for MDMO-PPV:PCBM \cite{dyakonov2002}.Two different temperature ranges can be observed. In the high temperature regime (HTR) between around 150 to 300~$K$, the open circuit voltage is decreasing with increasing temperature as well as the extracted charge carrier density. The low temperature regime (LTR), ranging from 50 to 150~$K$, shows an increase of $n$ stored in the device with raising temperature and a saturation effect for the $V_{oc}$ at high light intensities. For lower light intensities an increase of open circuit voltage with increasing temperature is observed. The illumination level was varied more than two orders of magnitude from 0.01 to 3.2 $suns$ leading to an increase of both $n$ and  $V_{oc}$. At 300~$K$ the extracted charge carrier density increases only by a factor of $\sim$~10 despite the more than 300-fold increase in illumination. This nonlinear behavior can be explained by the dependence of the recombination rate on the charge carrier density $R=k n^{\lambda+1}$, where $\lambda$ was shown to be in the range of 1.75 for P3HT:PCBM solar cells at room temperature \cite{foertig2009}.
\begin{figure}[tbp] 
   \centering
   \includegraphics[scale=0.5]{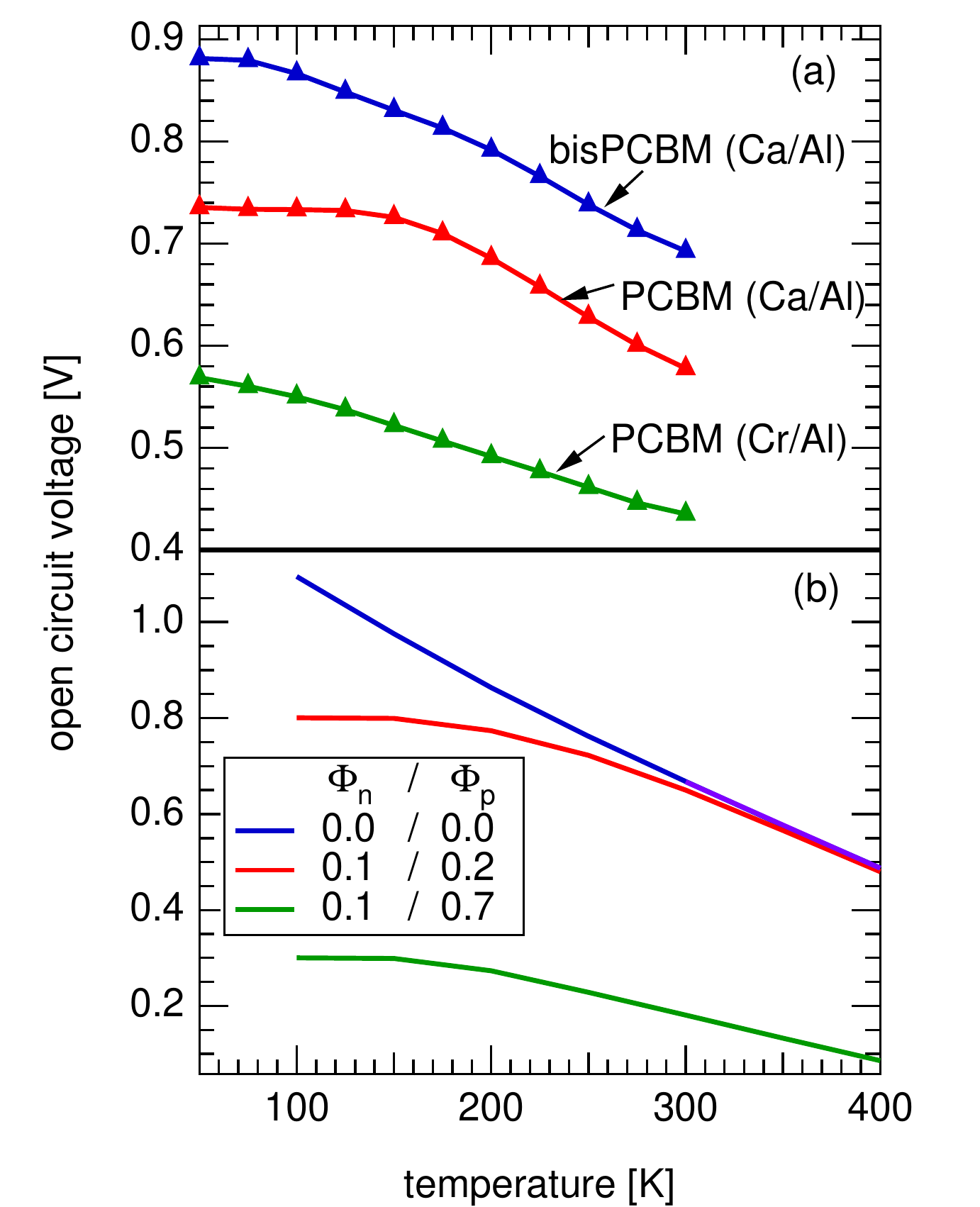} 
   \caption{(Color online) (a) Temperature dependent $V_{oc}$ behavior for three different solar cells. PCBM and bisPCBM were used as acceptor in blends with P3HT, the cathodes used are indicated in brackets. (b)  Temperature dependent $V_{oc}$ behavior for different energy barrier heights at the contacts obtained by macroscopic simulation.}
   \label{fig:Fig2}
\end{figure}
The temperature dependent open circuit voltages for all investigated solar cells at 1 sun are shown in Fig.~\ref{fig:Fig2} (a). The P3HT:bisPCBM has an overall higher $V_{oc}$ than the P3HT:PCBM solar cell with the same Ca/Al contact, which is due to the higher LUMO level of bisPCBM compared to PCBM \cite{lenes2008}. The saturation effect in the LTR is also visible, although it is not as pronounced. The P3HT:PCBM solar cell with Cr/Al contact has a lower open circuit voltage than the cell with the Ca/Al contact, indicating that in this case the contact is limiting, since the bulk properties have not changed. The temperature dependence of $n$ for the P3HT:bisPCBM and the Cr/Al P3HT:PCBM cell show the same behavior as  the Ca/Al P3HT:PCBM cell, but with slightly lower values (not shown).

Investigating the influence of the charge carrier density generated in the cell on the open circuit voltage, the $V_{oc}(T)$ was fitted using Eqn.~(\ref{eq:voc2}) under the assumption of $n=p$. $n(T)$ obtained from CE measurements provides $E_g$ and $N_c$ as fitting parameters. For the fits shown in Fig.~\ref{fig:Fig1}(a) (dotted lines) only the HTR was used, where Eqn.~(\ref{eq:voc2}) describes the $V_{oc}$ behavior very well. Expanding the fit using $E_g$ and $N_c$ from the HTR to the LTR shows a clear discrepancy between the fit and the measured voltage. This implies that the polaron concentration is not responsible for the saturation of $V_{oc}$ at low temperatures, although it can explain the HTR range well. We propose that the saturation effect at low temperatures is not caused by bulk properties but by the contacts,  limiting the maximum achievable voltage.

To support this proposition macroscopic simulations of temperature dependent IV--characteristics were performed. There, Poisson, continuity and drift--diffusion equations were solved simultaneously by an iterative approach explained in Ref.~\cite{wagenpfahl2010, deibel2008b} in more details. The mobilities of electrons and holes were assumed to be balanced, their temperature dependences were calculated by the Gaussian disorder model  \cite{bassler1993} by $ \mu(T)=\mu_{0}\exp(-(2\sigma/3k_{B}T)^2)$, with $\mu_0=1.1 \times 10^{-7}$m$^2$V$^{-1}$s$^{-1}$ and $\sigma$=0.06~eV as width of the Gaussian density of states. Recombination was considered by the reduced Langevin model, whereas field and temperature dependent polaron pair dissociation was not taken into account. The other parameters used are tabulated in Ref.~\cite{wagenpfahl2010}. In addition, different energy barrier heights for charge injection into the blend for electrons $\Phi_n$ and holes $\Phi_p$ were considered in order  to investigate their influence on the temperature dependence of the simulated open circuit voltage.  

The results are depicted in Fig. \ref{fig:Fig2} (b). Without an injection barrier, implying \textit{perfect} ohmic contacts at both electrodes, a linear temperature dependence of $V_{oc}$ over the whole temperature range from 100 to 400~$K$ is determined. A linear fit of this data leads to an intersection at T=0~$K$ about 1.2~$eV$ which is above the given $E_g$ of 1.1~$eV$. Based on the square root shaped density of states as necessary approximation for the numerical calculations, this effect originates from the fixed boundary conditions of thermionic emission at the contacts and charge carrier densities above the effective density of states in the bulk. Including a barrier for electrons of 0.1~$eV$ and holes of 0.2~$eV$ already shows a saturation effect of $V_{oc}$ at lower temperatures at 0.8~V. For higher temperatures these relatively low barrier heights can still be considered as ohmic contacts, reducing the open circuit voltage only slightly. Raising the sum of the electron and hole injection barrier to 0.8~$eV$ shows a high impact on $V_{oc}$ even at high temperatures. The difference in $V_{oc}$ compared to the cell with  ohmic contacts of 0.53~$V$ at 300~$K$ implies that non ohmic contacts limit the open circuit voltage at room temperature. Furthermore, it can be seen that it is not appropriate to extrapolate the linear $V_{oc}$ range to T=0~$K$ and use this intersection point as $E_g/q$, as long as the contacts barriers are not zero.

The simulation explains the saturation behavior of $V_{oc}$ at low temperatures and can also predict the value at which it occurs, namely  $qV_{oc}=E_{g}-(\Phi_n+\Phi_p)$. For example in the simulation with an injection barrier for the electrons of 0.1~$eV$ and for the holes of 0.7~$eV$, together with $E_g$=1.1~eV, the saturated $V_{oc}$ is 0.3~$V$. This makes it possible to determine the sum of the barriers by temperature dependent IV measurements and determine the built in voltage $V_{bi}$, which is the value at which $V_{oc}$ saturates. 
 
Comparing the results of the simulation with the experiment shows qualitatively good agreement. In both solar cells with Ca/Al contact the saturation of the open circuit voltage can be observed at low temperatures. If the work function of the Ca/Al cathode was the same for these two cells, implying the same $V_{bi}$, both cells should saturate at same voltage. We propose that the difference of $\sim$ 0.1~V in the saturation voltage can be explained by different interface dipoles between the PCBM/Ca and bisPCBM/Ca interface, thus changing the corresponding work functions. Indeed, interface dipoles have often been observed for interfaces of evaporated small molecules to metals~\cite{hill1998} and polymer:fullerene blends to metals \cite{guan2010}, and it is a credible assumption that different material combinations will influence the magnitude of the dipole. Thus, within our scenario, $V_{oc}$ in the LTR is limited by the contacts and only slightly influenced by bulk effects such as the charge generation (see light intensity dependence in Fig.~\ref{fig:Fig1}a).
   
The difference of $V_{oc}$ in the HTR is mainly caused by the difference of the LUMO levels of PCBM and bisPCBM, as the charge density is in the same range. This indicates that in this temperature range the open circuit voltage is mainly affected by bulk properties such as the photogeneration an recombination of polarons. Thus, the equation given by Brabec et al.~\cite{brabec2001}, $V_{oc}\approx\left|\text{HOMO}_{\text{donor}}-\text{LUMO}_{\text{acceptor}}\right| - 0.3$~V, can be used as a rule of thumb.

Simulation and experiment are also consistent for the cells with a limiting contact, as experimentally demonstrated by using a Cr/Al cathode, because of the low workfunction of Cr creating a non ohmic contact. In this case, the open circuit voltage is limited even in the HTR, in contrast to the P3HT:PCBM cell with the Ca/Al cathode, although the bulk properties are the same. Such a device is completely contact limited, despite the $V_{oc}$ slightly increases with lower temperature. Thus we conclude, in the case of limiting contacts, $V_{oc}$ is determined by MIM model \cite{mihailetchi2003}.

In conclusion, IV and CE measurements of organic BHJ solar cells were performed at various temperatures and illumination intensities to investigate the relation between the open circuit voltage and corresponding charge carrier density. A linear temperature dependence of $V_{oc}$ at higher temperatures was observed. The saturation at lower temperatures was indentified to be caused by injection barriers for charge carriers at the contact, as verified by macroscopic simulations. If these barriers are high and lead to non ohmic contacts, they can also reduce $V_{oc}$ and therefore the solar cell efficiency at working conditions i.e. room temperature. 

D.R.'s work is financed by the European Commission in the framework of the Dephotex Project. C.D. gratefully acknowledges the support of the Bavarian Academy of Sciences and Humanities.  V.D.'s work at the ZAE Bayern is financed by the Bavarian Ministry of Economic Affairs, Infrastructure, Transport and Technology.

\end{document}